# Electron Attachment Induced Shape Resonances in AT Base Pairs

Sneha Arora[a], Jishnu Narayanan S J[a] and Achintya Kumar Dutta[a*]

[a] *Department of Chemistry, Indian Institute of Technology Bombay, Powai, Mumbai 400076, India.*

**Abstract**

In this work, we investigated the influence of base pairing and π-π stacking interactions on electron attachment induced shape resonances in the adenine-thymine (AT) base pair. Resonance positions and widths are computed using a DLPNO based equation of motion coupled-cluster approach in conjunction with the Padé analytical continuation method. Seven π* shape resonances are identified for both linear and stacked AT geometries, consistent with the total number of resonances in isolated adenine and thymine. Natural orbital analysis reveals that low-energy resonances exhibit significant electron density delocalization over both nucleobases. This delocalization is enhanced in the stacked geometry, leading to appreciable stabilization and increased lifetimes of the resonance states. These results highlight the important role of intermolecular interactions in modulating electron attachment processes in DNA.

*achintya@chem.iitb.ac.in

## 1. Introduction

The study of electron interactions with biomolecules, particularly DNA, has attracted significant attention in recent decades due to their important role in radiation induced damage and associated biological effects.[1] Ionizing radiation, such as X-rays or γ-rays, generates a large number of secondary electrons (SEs),[2–5] along with other products.[2,6–10] These SEs play an important role in radiative cellular damage, including mutations, disruption of genomic stability, and apoptosis.[2–5,11] The energy distribution of SEs typically peaks around 9 eV,[12] with most electrons having energies below 20 eV.[3,13–17] These electrons undergo successive inelastic collisions, producing low-energy electrons (LEEs) with a spectrum that peaks below 1 eV and a half-width of approximately 2 eV.[13–19] Previous studies have shown that LEEs with energies below the ionization threshold of DNA (~7.5-10 eV)[20,21] can induce significant molecular damage. Notably, Woldhuis et al.[22] and Boudaïffa et al.[23] demonstrated that LEEs can directly interact with DNA in aqueous environments, leading to both single- and double-strand breaks.

Interaction of LEEs with biomolecules leads to the formation of transient negative ions (TNIs*) or resonance states with lifetimes ranging from $10^{-12}$ to $10^{-15}$ seconds.[24] These states may undergo autodetachment, relax to bound radical anions, or dissociate via dissociative electron attachment (DEA),[25–28] resulting in bond cleavage.[23,29–37] Resonance states play a crucial role across diverse domains, from high energy processes such as plasma dynamics and attosecond spectroscopy,[38,39] to low energy electron-molecule interactions relevant to interstellar chemistry and radiation damage in biomolecules. They are generally classified as one-particle (1p) or two-particle one-hole (2p1h) resonances based on the electronic configuration of the neutral target prior to electron attachment.[35,36,40–45] A one-particle resonance, also called a shape resonance, is formed when an incoming electron gets temporarily trapped in the interaction potential with a neutral molecule. The shape of the potential is formed by the combined effects of the attractive polarization potential and the repulsive centrifugal potential.[46] However, electron attachment to an excited state leads to a 2p1h resonance state which is characterized by a hole in one of the occupied molecular orbitals.[35]

Resonance states are non-stationary, finite lifetime states that couple to the continuum and involve a superposition of multiple configurations.[47–56] These states are characterized by Siegert energies, $E_{res} = E_R - i\Gamma/2$, where the real part denotes the resonance energy and the imaginary part is related to the inverse lifetime.[47] These states exhibit bound-state like behavior within the interaction region but possess an outgoing oscillatory tail outside the potential barrier, rendering the associated wavefunction non-square integrable. Consequently, conventional bound state electronic structure methods are not directly applicable, necessitating specialized approaches for their description. The development of so-called $L^2$ based *ab initio* methods, including complex scaling (CS),[48–52] complex absorbing potential (CAP),[52–60] and stabilization approach,[61–64] has significantly advanced the theoretical treatment of resonances in many-electron systems. Among these, the stabilization approach is particularly appealing because it enables the use of standard quantum chemical methods without modification.[61–64]

Electron attachment induced resonances in isolated DNA and RNA nucleobases have been extensively studied using both experimental and theoretical approaches.[34–36,42,45,65–75] Scattering calculations have been extensively used to identify resonances in nucleobases, nucleosides, and nucleotides, mostly using DFT as the method of calculation.[19] The orbital stabilization method has also been applied to isolated nucleobases within a DFT framework using Koopmans' theorem.[53] We have recently demonstrated that TD-DFT method can provide a low-cost way to simulate resonances of nucleobases,[76] if proper exchange-correlation functionals are used. Accurate electron correlation methods are also used for resonance calculations. For example, shape resonances have been investigated using the CAP/SAC-CI approach,[53] while Matsika and co-workers combined the stabilization method with EOM-EA-CCSD and XMCQDPT2 approaches to predict both shape and core-excited resonances in nucleobases.[35,77,78]

Most theoretical studies of anionic resonances in genetic material have focused on isolated nucleobases as model systems. However, recent studies indicate that environmental effects, including aqueous surroundings and amino acids, can significantly influence nucleobase resonances.[74,79] Since DNA is inherently double stranded, it is essential to investigate electron attachment in hydrogen bonded base pairs, which serve as more realistic models of the genetic environment. Base pairs capture key intermolecular interactions and provide direct insight into how low-energy electrons interact with and potentially damage DNA. Early *ab initio* studies by Sevilla and co-workers[80] examined electron-attached states in AT and GC base pairs at the Hartree Fock level, revealing the formation of valence-type radical anions capable of inducing inter-base proton transfer. Subsequent studies by Adamowicz and co-workers[81] reported a negative adiabatic electron affinity (−0.40 eV) for the AT base pair, indicating unfavorable electron attachment, whereas the GC base pair supports both dipole-bound and valence-bound states.[82] Despite these efforts, a systematic and quantitatively reliable characterization of electron attachment induced shape resonances in AT base pairs, particularly accounting for both base pairing and π-π stacking effects at a correlated wavefunction level, remains lacking. In this work, we address this gap by employing the EA-EOM-DLPNO-CCSD method[83,84] in combination with the Padé-based stabilization approach[62–64,85] to characterize resonance energies, widths, and orbital nature in both linear and stacked AT geometries. This enables us to provide new insights into how base pairing and stacking modulate stability, delocalization, and lifetimes of resonance states in DNA model systems.

## 2. Computational Details:

The neutral ground state geometries of isolated nucleobases and base pairs were optimized at the RI-MP2/def2-TZVP[86,8788] level of theory. Multiple conformers of the adenine-thymine (AT) base pair, including stacked and hydrogen-bonded (linear) structures, were generated using the CREST program.[89] The lowest energy conformers for both stacked and linear AT configurations were subsequently re-optimized at the RI-MP2/def2-TZVP level and used for all further calculations. Vertical electron attachment energies were computed using the EA-EOM-DLPNO-CCSD[83,84] method with the NORMALPNO setting. The use of the DLPNO approximation enables the treatment of larger systems such as base pairs at a reduced computational cost compared to canonical EOM-CCSD, while retaining a reliable description of electron-attached states.[90,91] This is particularly important for investigating base pairs, where system size becomes a limiting factor for conventional correlated methods. Resonance stabilization curves were constructed using the cc-pVDZ basis set[92] augmented additional diffuse function of 2s, 2p, and 2d angular momentum type to all heavy atoms present in the molecule following reference.[71] The expanded basis set is denoted by the notation cc-pVDZ+2s2p2d. The relevant auxiliary basis set was selected using Autoaux[93] tool of ORCA 5.0.3 software.[94,95] The optimized structures of isolated nucleobases and AT (stacked and linear) base pairs are provided in the Supplementary Information. (Figure S1) The resonance via padé (RVP) approach[62] was used to extract complex resonance energies (positions and widths) from the stable regions of the stabilization curves. All RVP calculations were performed using the open source Automatic RVP program.[96] The molecular systems investigated include thymine (T), adenine (A), the adenine-thymine base pair (AT), and their stacked counterparts, denoted as sTT, sAA, and sAT.

## 3. Results and Discussion:

In this work, we focus exclusively on shape resonances. To assess the effect of base pairing on the positions and widths of these resonances, it is essential to first characterize the shape resonances of the isolated nucleobases. This serves as a benchmark for evaluating the accuracy of the present theoretical approach, as reliable reference data are available for individual nucleobases. [34–36,42,45,65–75,97]

### 3.1 Thymine shape resonances:

The shape resonance states of thymine have been experimentally characterized using low-energy electron transmission spectroscopy (ETS).[98] Three resonance states were identified, which are blue-shifted relative to those of uracil by approximately 0.07, 0.13, and 0.22 eV. This shift is attributed to the σ-electron donating effect of the C5 methyl substituent in thymine.[98]

In the present work, we identify three low-lying π* shape resonances of the thymine anion using the EA-EOM-DLPNO-CCSD method in combination with the RVP approach. These states are labeled as 1π*, 2π*, and 3π* in order of increasing energy. Figure 1A shows the corresponding stabilization plot along with the natural orbitals associated with each resonance state. Table 1 summarizes the computed resonance positions and widths obtained using the cc-pVDZ+2s2p2d basis set. The three π* resonances are located at 0.69, 2.35, and 5.69 eV, respectively. The lowest lying 1π* state has a width of 0.014 eV, corresponding to a lifetime of approximately 47 fs. The 2π* state is less stable, with a shorter lifetime of ~24 fs, while the 3π* state appears at higher energy and exhibits the largest width, corresponding to a lifetime of ~6 fs. Overall, the increase in resonance energy is accompanied by broader widths and shorter lifetimes, indicating stronger coupling with the continuum.

The calculated positions of the 1π* and 2π* resonances are in very good agreement with GPA-EA-EOM-CCSD results reported by Fennimore and Matsika[36] (See Table 1). However, the energy of the 3π* resonance is overestimated by approximately 0.67 eV. High-lying resonances are known to exhibit stronger coupling with nearby Feshbach states, making them more challenging to describe accurately within a single-reference framework such as EA-EOM-CCSD. In comparison with CAP/SAC-CI results reported by Sommerfield and Ehara,[53] the 3π* resonance energy is underestimated by ~0.55 eV, while the lower-lying resonances are well reproduced. R-matrix calculations show a wider variation in resonance energies, likely reflecting differences in the underlying electronic structure methods.[68,99]

The resonance widths obtained in this work are consistently smaller than those reported in previous studies (Table 1). However, significant variation in reported widths exists in the literature, with a general consensus that the width increases with resonance energy. Notably, other theoretical approaches predict larger widths for the 3π* state, typically in the range of 0.4 to 1.0 eV, whereas the present results yield a comparatively smaller value. The absence of experimental data for resonance widths makes direct validation difficult. It is important to note that resonance widths are extremely sensitive to the choice of electronic structure method, the $L^2$ technique used to extract complex energies, and the basis set employed. Consequently, variations in these factors can lead to significant discrepancies across different studies, with widths being particularly sensitive compared to resonance energies. Despite the quantitative differences, the qualitative trends in resonance widths and their dependence on energy are consistent with previous theoretical studies.

### 3.2 Adenine shape resonances:

Adenine is known to have one of the lowest electron affinities among the DNA nucleobases.[100] The valence-bound anionic states of isolated adenine are neither vertically nor adiabatically stable, while its dipole-bound states are only weakly bound due to the relatively small dipole moment.[101] Consequently, shape resonance states play a dominant role in the electron capture process for adenine. In the present work, we characterized four low-lying π* shape resonances of the adenine anion. The corresponding stabilization plot and natural orbitals are shown in Figure 1B, and the computed resonance energies and widths are summarized in Table 2 along with comparisons to previous theoretical and experimental results. The four resonances are located at 1.11, 1.98, 2.97, and 7.20 eV, respectively. Experimentally, only three shape resonances have been reported in electron transmission spectroscopy studies.[98] The first three calculated resonance energies differ from experiment by approximately 0.5 to 0.8 eV, similar to the deviations observed for thymine. Such deviations are typical for stabilization based approaches combined with finite basis sets and have been reported in previous studies.[35] The fourth resonance around 7.20 eV lies outside the experimentally reported range and may not have been observed due to its weak intensity or broad character.

The 1π* and 2π* resonances exhibit relatively narrow widths, corresponding to lifetimes of approximately 50 fs and 25 fs, respectively. The 3π* resonance is broader, with a width of 0.038 eV and a corresponding lifetime of ~17 fs. The highest-lying 4π* resonance is the broadest and shortest-lived, with a lifetime of ~3 fs. This systematic decrease in lifetime with increasing resonance energy reflects stronger coupling with the continuum for higher-lying states. The positions of the first three π* resonances are in good agreement with the results reported by Fennimore and Matsika,[36] using the analytically continued GPA-EA-EOM-CCSD method. However, the energy of the 4π* resonance is overestimated by approximately 0.4 eV. This discrepancy may arise from the increased diffuseness and possible mixing of the 4π* state with nearby Feshbach resonances, which makes it more challenging to describe within a single-reference framework.

The calculated resonance widths are generally smaller than those obtained using the GPA approach, except for the 4π* state. These differences may originate from methodological variations between the RVP and GPA approaches, as well as from differences in basis set treatment and the sensitivity of width calculations to the stabilization procedure. In addition, the possible mixing of the 4π* resonance with nearby Feshbach states may further complicate its description.

Despite these quantitative differences, the overall trends in resonance widths and lifetimes are consistent with previous theoretical studies. The resonance energies are also in reasonable agreement with CAP/SAC-CI results reported by Sommerfeld and Ehara.[53]

### 3.3 AT base pair (linear):

We identify seven π* shape resonances for the AT base-pair anion, consistent with the combined number of resonances observed in the isolated adenine and thymine nucleobases. The stabilization plot obtained at the RVP/EA-EOM-DLPNO-CCSD/cc-pVDZ+2s2p2d level of theory is shown in Figure 3A. To confirm the nature of these resonances, we analyze the corresponding natural orbitals in the stabilized regions of the plot. Figure 2 presents the natural orbitals of all seven π* resonances in the AT base pair, along with those of the isolated nucleobases for comparison. The lowest lying 1π* resonance is located at 0.67 eV with a width of 0.011 eV, while the highest 7π* resonance appears at 6.80 eV with a width of 0.116 eV, illustrating the broad range of resonance energies and lifetimes induced by base pairing.

To assess the effect of base pairing, we compare the resonance energies and widths of the AT base-pair anion with those of the isolated adenine and thymine anions. The results are summarized in Table 3, where resonance energies and widths are extracted from the well-defined stable regions of the stabilization curves. The comparison shows that the 1π*, 4π*, and 6π* resonances of AT are predominantly thymine-centered, whereas the 2π*, 3π*, 5π*, and 7π* resonances are primarily adenine-centered. This assignment is supported by both the spatial character of the natural orbitals and the similarity in resonance energies and widths relative to the corresponding isolated nucleobase states. Notably, the lowest four π* resonances exhibit significant delocalization of electron density over both nucleobases, indicating that intermolecular interactions between adenine and thymine play a key role in shaping the electronic structure of the anionic states. While simulating the TNIs of AT base pair, we observe significant interactions between the wave-function of adenine and thymine. As evident from Figure 2, AT-1π* resonance is thymine centered with an energy of 0.67 eV. It is slightly stabilized compared to the corresponding thymine 1π*, which has an energy of 0.69 eV. Moreover, it also has a contribution from the 1π* resonance of adenine which becomes clear in the natural orbital plot. The second AT-2π* state is primarily adenine-centered with an energy of 1.26 eV and shows contribution from thymine 1π* state. It is destabilized by the 0.15 eV compared to the corresponding adenine 1π* resonance at 1.11 eV. Similarly, the adenine-centered AT-3π* resonance, which is similar to adenine 2π* state, also has a contribution from thymine. AT-3π* state is also energetically stabilized by 0.05 eV compared to its corresponding adenine 2π* shape resonance. The AT-4π* resonance is thymine-centered and is located at 2.43 eV, and its natural orbital plot is similar to the thymine 2π* state, but with slight contribution from adenine orbital state. This state is destabilized by 0.08 eV relative to the isolated thymine 2π* resonance.

For the higher-lying resonances, the AT-5π* state is largely localized on adenine but still exhibits contributions from thymine orbitals, indicating residual intermolecular coupling. The AT-6π* and AT-7π* resonances are primarily thymine- and adenine-centered, respectively. The trends observed in AT base pair show contrast to those observed in GC base pair in two respects.[102] First, all the low-lying resonances observed in adenine and thymine are identifiable in AT base pair. Although their positions got shifted, there is no qualitative change in their character. This behavior contrasts with that observed in GC base pairs, where certain low-lying resonances are absent. Second, in case of GC base pair, all the purine centered resonances are blue-shifted, and pyrimidine centered resonances are red-shifted. On the contrary, the 2π* state of thymine (4π* states in AT) gets blue-shifted, and the 2π* and 4π* states of adenine (3π* and 7π* in AT) are red-shifted.

The width of the resonance states in AT base pair generally follow trends similar to those of the isolated nucleobases. Resonances that undergo red-shifts upon base pairing tend to exhibit reduced widths, corresponding to longer lifetimes, whereas blue-shifted resonances typically show increased widths and shorter lifetimes.

Overall, several resonances become narrower upon formation of the base pair, indicating enhanced stabilization due to intermolecular interactions. However, this effect is not uniform across all states, suggesting that the influence of base pairing is strongly state-dependent and governed by the extent of orbital mixing and spatial localization.

**3.4 Stacking vs base pair effect:**

In biological systems, base pairs are arranged in a stacked configuration within the DNA double helix, giving rise to π-π interactions between adjacent bases. These interactions can significantly influence the stability and decay dynamics of resonance states. To assess the impact of stacking, we performed resonance calculations for the stacked AT (sAT) configuration and compared the results with those of the linear (hydrogen bonded) AT base pair.

Seven π* shape resonances are identified for the sAT anion, consistent with the linear AT case. The corresponding stabilization plot is shown in Figure 3B.

The resonance energies and widths for both stacked and linear AT configurations are summarized in Table 4. The lowest-lying sAT-1π* resonance shows a slight stabilization relative to the linear AT case, with a red-shift of 0.03 eV and a small decrease in width (~0.002 eV), indicating a longer lifetime. A more pronounced effect is observed for the sAT-2π* state, which appears at 0.88 eV and exhibits a significant red-shift of 0.38 eV compared to its linear counterpart, along with an increase in lifetime of approximately 6 fs. Similar trends observed for higher-lying resonances, where both the resonance energies and widths are generally exhibit red-shifts in the stacked configuration relative to the linear AT base pair. The enhanced stabilization observed in the stacked configuration can be attributed to stronger π-π interactions between the nucleobases. In the stacked geometry, the head-on overlap of π* orbitals facilitate greater delocalization of the excess electron compared to the side-on interactions present in the linear base pair. This increased delocalization presumably leads to lower resonance energies and longer lifetimes, indicating more stable resonance states in the stacked configuration.

All resonance states in the stacked AT configuration are red-shifted relative to those in the isolated nucleobases. The magnitude of the shift varies across different states, ranging from ~0.05 eV for the lowest 1π* resonance to as much as ~0.7 eV for the higher lying 7π* state. This variation indicates that the effect of stacking is strongly state-dependent.

The increased lifetimes observed in stacked configurations suggest a higher probability for subsequent nuclear relaxation, which may facilitate the formation of stable anionic states and influence damage pathways in DNA.

### 3.5 The effect of base sequence:

Experimental studies have shown that the base sequence plays an important role in determining the extent of electron attachment induced DNA damage.[103] To further investigate this effect, we examine stacked adenine-adenine (sAA) and thymine-thymine (sTT) base pairs, which allow us to isolate the role of homobase stacking. This provides additional insight into how stacking interactions influence the stability and character of anionic resonance states. The corresponding stabilization plots are provided in Figure S2.

For the stacked AA system, we identified eight $\pi^*$ shape resonances, which are approximately twice the number observed for the isolated adenine nucleobase. The corresponding natural orbitals are shown in Figure 5. This increase arises from orbital coupling between the two adenine units, leading to the formation of delocalized $\pi^*$ states across the dimer. A similar trend is observed for the stacked TT system, where six $\pi^*$ resonances are identified (Figure 6), compared to three for the isolated thymine nucleobase. In both cases, interaction between equivalent $\pi^*$ orbitals of the two nucleobases result in splitting into stabilized (red-shifted) and destabilized (blue-shifted) resonance states.

The resonance states that undergo red-shifts relative to the isolated nucleobases also exhibit reduced widths, corresponding to longer lifetimes, whereas blue-shifted states show increased widths and shorter lifetimes (Table 5). This behavior is consistent with the trends observed for the AT base pair and reflects the relationship between resonance stabilization and coupling to the continuum. However, the overall stabilization observed in sAA and sTT is significantly smaller than that found in the stacked AT system, indicating that heterobase stacking leads to stronger intermolecular interactions and greater stabilization of resonance states compared to homobase stacking. These results suggest that base sequence can modulate the efficiency of electron trapping and redistribution in DNA, with heterobase stacking providing more favorable pathways for charge delocalization.

## 4. Conclusions

In this work, we investigated the effect of base pairing and π-π stacking on electron attachment induced shape resonances in adenine-thymine (AT) base pairs using the EA-EOM-DLPNO-CCSD method combined with the Padé-based stabilization approach.

For the linear AT base pair, we identified seven π* shape resonances, corresponding to the combined resonances of the isolated adenine and thymine nucleobases. While the overall character of these resonances remains largely preserved upon base pairing, their energies and lifetimes are significantly modulated due to intermolecular interactions. In particular, several low-lying resonances exhibit partial delocalization over both nucleobases, accompanied by shifts in resonance positions and widths that reflect changes in stability and coupling to the continuum. Stacking interactions further enhances these effects. In the stacked AT configuration, all resonance states exhibit overall stabilization (red-shifts) and increased lifetimes relative to both the linear base pair and isolated nucleobases. This behavior is attributed to stronger π-π interactions and enhanced delocalization of the excess electron. In contrast, homobase stacking (AA and TT) shows comparatively weaker stabilization, highlighting the importance of heterobase interactions in modulating resonance properties.

These results demonstrate that intermolecular interactions in DNA, particularly base pairing and stacking, play a crucial role in determining the stability, lifetime, and spatial character of electron attached states. The increased stabilization and lifetime of resonance states in stacked configurations suggest a higher likelihood of electron trapping and subsequent chemical processes, which may influence pathways of radiation induced damage in DNA. Future studies incorporating more realistic models, including the sugar-phosphate backbone and environmental effects, will be essential for a comprehensive understanding of electron induced processes in DNA. Work is in progress towards that direction.

**Supplementary material:** The optimized geometries of isolated nucleobases, base-pair and stacked nucleobases, additional stabilization plots are provided in the Supporting Information.

**Acknowledgments**

The authors acknowledge financial support from the IIT Bombay, UGC-India and the Prime Minister's Research Fellowship. IIT Bombay super computational facility, and C-DAC Supercomputing resources (Param Smriti, Param Brahma) for computational time. AKD acknowledges the research fellowship funded by the EU NextGenerationEU through the Recovery and Resilience Plan for Slovakia under project No. 09I03-03-V04-00117.

**References:**

(1) Rezaee, M.; Adhikary, A. The Effects of Particle LET and Fluence on the Complexity and Frequency of Clustered DNA Damage. *DNA* **2024**, *4* (1), 34–51.

(2) O'Neill, P. Radiation-Induced Damage in DNA. In *Studies in Physical and Theoretical Chemistry*; Jonah, C. D., Rao, B. S. M., Eds.; Radiation Chemistry; Elsevier, 2001; Vol. 87, pp 585–622.

(3) Alizadeh, E.; Sanche, L. Precursors of Solvated Electrons in Radiobiological Physics and Chemistry. *Chem. Rev.* **2012**, *112* (11), 5578–5602.

(4) Narayanan S J, J.; Tripathi, D.; Dutta, A. K. Doorway Mechanism for Electron Attachment Induced DNA Strand Breaks. *J. Phys. Chem. Lett.* **2021**, *12* (42), 10380–10387.

(5) Narayanan S J, J.; Tripathi, D.; Verma, P.; Adhikary, A.; Dutta, A. K. Secondary Electron Attachment-Induced Radiation Damage to Genetic Materials. *ACS Omega* **2023**, *8* (12), 10669–10689.

(6) Cadet, J.; Bellon, S.; Douki, T.; Frelon, S.; Gasparutto, D.; Muller, E.; Pouget, J.-P.; Ravanat, J.-L.; Romieu, A.; Sauvaigo, S. Radiation-Induced DNA Damage: Formation, Measurement, and Biochemical Features. **2004**, *23* (1), 12.

(7) Berthel, E.; Ferlazzo, M. L.; Devic, C.; Bourguignon, M.; Foray, N. What Does the History of Research on the Repair of DNA Double-Strand Breaks Tell Us?—A Comprehensive Review of Human Radiosensitivity. *Int. J. Mol. Sci.* **2019**, *20* (21).

(8) Obodovskiy, I. Radiation Therapy. In *Radiation*; Elsevier, 2019; pp 387–396.

(9) *DNA Damage, DNA Repair and Disease: Volume 1*; Dizdaroglu, M., Lloyd, R. S., Dizdaroglu, M., LLoyd, R. S., Eds.; The Royal Society of Chemistry, 2020.

(10) *DNA Damage, DNA Repair and Disease: Volume 2*; Dizdaroglu, M., Lloyd, R. S., Dizdaroglu, M., LLoyd, R. S., Eds.; The Royal Society of Chemistry, 2020.

(11) Becker, D.; Kumar, A.; Adhikary, A.; Sevilla, M. D. Gamma- and Ion-Beam DNA Radiation Damage: Theory and Experiment. **2020**, 426–457.

(12) Pimblott, S. M.; LaVerne, J. A. Production of Low-Energy Electrons by Ionizing Radiation. *Proc. 11th Tihany Symp. Radiat. Chem.* **2007**, *76* (8), 1244–1247.

(13) Herbert, J. M.; Coons, M. P. The Hydrated Electron. *Annu. Rev. Phys. Chem.* **2017**, *68* (1), 447–472.

(14) Walker, D. C. The Hydrated Electron. *Q. Rev. Chem. Soc.* **1967**, *21* (1), 79–108.

(15) Steenken, S. Purine Bases, Nucleosides, and Nucleotides: Aqueous Solution Redox Chemistry and Transformation Reactions of Their Radical Cations and E~ and OH Adducts. *Chem. Rev.* **1989**, *89* (3), 503–520.

(16) Kumar, A.; Sevilla, M. D. Low-Energy Electron (LEE)-Induced DNA Damage: Theoretical Approaches to Modeling Experiment. In *Handbook of Computational Chemistry*; Leszczynski, J., Ed.; Springer Netherlands: Dordrecht, 2012; pp 1215–1256.

(17) Ma, J.; Kumar, A.; Muroya, Y.; Yamashita, S.; Sakurai, T.; Denisov, S. A.; Sevilla, M. D.; Adhikary, A.; Seki, S.; Mostafavi, M. Observation of Dissociative Quasi-Free Electron Attachment to Nucleoside via Excited Anion Radical in Solution. *Nat. Commun.* **2019**, *10* (1), 102.

(18) Cobut, V.; Frongillo, Y.; Patau, J. P.; Goulet, T.; Fraser, M.-J.; Jay-Gerin, J.-P. Monte Carlo Simulation of Fast Electron and Proton Tracks in Liquid Water – I. Physical and Physicochemical Aspects. *Radiat. Phys. Chem.* **1998**, *51*, 229–243.

(19) Kumar, A.; Sevilla, M. D.; Sanche, L. How a Single 5 eV Electron Can Induce Double-Strand Breaks in DNA: A Time-Dependent Density Functional Theory Study. *J. Phys. Chem. B* **2024**, *128* (17), 4053–4062.

(20) Pluhařová, E.; Schroeder, C.; Seidel, R.; Bradforth, S. E.; Winter, B.; Faubel, M.; Slavíček, P.; Jungwirth, P. Unexpectedly Small Effect of the DNA Environment on Vertical Ionization Energies of Aqueous Nucleobases. *J. Phys. Chem. Lett.* **2013**, *4* (21), 3766–3769.

(21) Pluhařová, E.; Slavíček, P.; Jungwirth, P. Modeling Photoionization of Aqueous DNA and Its Components. *Acc. Chem. Res.* **2015**, *48* (5), 1209–1217.

(22) Woldhuis, J.; Verberne, J. B.; Lafleur, M. V. M.; Retèl, J.; Blok, J.; Loman, H. γ-Rays Inactivate ϕX174 DNA in Frozen Anoxic Solutions at −20°C Mainly by Reactions of Dry Electrons. *Int. J. Radiat. Biol. Relat. Stud. Phys. Chem. Med.* **1984**, *46* (4), 329–330.

(23) Boudaïffa, B.; Cloutier, P.; Hunting, D.; Huels, M. A.; Sanche, L. Resonant Formation of DNA Strand Breaks by Low-Energy (3 to 20 eV) Electrons. *Science* **2000**, *287* (5458), 1658–1660.

(24) Fabrikant, I. I.; Eden, S.; Mason, N. J.; Fedor, J. *Recent Progress in Dissociative Electron Attachment: From Diatomics to Biomolecules*, 1st ed.; Elsevier Inc., 2017; Vol. 66.

(25) Pan, X.; Cloutier, P.; Hunting, D.; Sanche, L. Dissociative Electron Attachment to DNA. *Phys. Rev. Lett.* **2003**, *90* (20), 208102.

(26) Wang, X.-D.; Xuan, C.-J.; Feng, W.-L.; Tian, S. X. Dissociative Electron Attachments to Ethanol and Acetaldehyde: A Combined Experimental and Simulation Study. *J. Chem. Phys.* **2015**, *142* (6), 064316.

(27) Ptasinska, S. A Missing Puzzle in Dissociative Electron Attachment to Biomolecules: The Detection of Radicals. *Atoms* **2021**, *9* (4).

(28) Bald, I.; Čurík, R.; Kopyra, J.; Tarana, M. Dissociative Electron Attachment to Biomolecules. In *Nanoscale Insights into Ion-Beam Cancer Therapy*; Solov'yov, A. V., Ed.; Springer International Publishing: Cham, 2017; pp 159–207.

(29) Kopyra, J. Low Energy Electron Attachment to the Nucleotide Deoxycytidine Monophosphate: Direct Evidence for the Molecular Mechanisms of Electron-Induced DNA Strand Breaks. *Phys. Chem. Chem. Phys.* **2012**, *14* (23), 8287–8289.

(30) Sanche, L. Low Energy Electron-Driven Damage in Biomolecules. *Eur. Phys. J. D* **2005**, *35* (2), 367–390.

(31) Gu, J.; Wang, J.; Leszczynski, J. Electron Attachment-Induced DNA Single Strand Breaks: C3′- O3′ σ-Bond Breaking of Pyrimidine Nucleotides Predominates. *J. Am. Chem. Soc.* **2006**, *128* (29), 9322–9323.

(32) Kumari, B.; Huwaidi, A.; Robert, G.; Cloutier, P.; Bass, A. D.; Sanche, L.; Wagner, J. R. Shape Resonances in DNA: Nucleobase Release, Reduction, and Dideoxynucleoside Products Induced by 1.3 to 2.3 eV Electrons. *J. Phys. Chem. B* **2022**, *126* (28), 5175–5184.

(33) Mukherjee, M.; Ragesh Kumar, T. P.; Ranković, M.; Nag, P.; Fedor, J.; Krylov, A. I. Spectroscopic Signatures of States in the Continuum Characterized by a Joint Experimental and Theoretical Study of Pyrrole. *J. Chem. Phys.* **2022**, *157* (20), 204305.

(34) Sieradzka, A.; Gorfinkiel, J. D. Theoretical Study of Resonance Formation in Microhydrated Molecules. II. Thymine-(H2O)n, n = 1,2,3,5. *J. Chem. Phys.* **2017**, *147* (3), 034303–034303.

(35) Fennimore, M. A.; Matsika, S. Core-Excited and Shape Resonances of Uracil. *Phys. Chem. Chem. Phys.* **2016**, *18* (44), 30536–30545.

(36) Fennimore, M. A.; Matsika, S. Electronic Resonances of Nucleobases Using Stabilization Methods. *J. Phys. Chem. A* **2018**, *122* (16), 4048–4057.

(37) Slaughter, D. S.; Rescigno, T. N. Breaking up Is Hard to Do. *Nat. Phys.* **2018**, *14* (2), 109–110.

(38) Ji, J.-B.; Guo, Z.; Driver, T.; Trevisan, C. S.; Cesar, D.; Cheng, X.; Duris, J.; Franz, P. L.; Glownia, J.; Gong, X.; Hammerland, D.; Han, M.; Heck, S.; Hoffmann, M.; Kamalov,

A.; Larsen, K. A.; Li, X.; Lin, M.-F.; Liu, Y.; McCurdy, C. W.; Obaid, R.; O'Neal, J. T.; Rescigno, T. N.; Robles, R. R.; Sudar, N.; Walter, P.; Wang, A. L.; Wang, J.; Wolf, T. J. A.; Zhang, Z.; Ueda, K.; Lucchese, R. R.; Marinelli, A.; Cryan, J. P.; Wörner, H. J. Attosecond X-Ray Core-Level Chronoscopy of Aromatic Molecules. *Phys. Rev. X* **2025**, *15* (4), 041031.

(39) Nisoli, M.; Decleva, P.; Calegari, F.; Palacios, A.; Martín, F. Attosecond Electron Dynamics in Molecules. *Chem. Rev.* **2017**, *117* (16), 10760–10825.

(40) Barrios, R.; Skurski, P.; Simons, J. Mechanism for Damage to DNA by Low-Energy Electrons. *J. Phys. Chem. B* **2002**, *106* (33), 7991–7994.

(41) Bryjko, L.; Mourik, T. van; Dora, A.; Tennyson, J. R-Matrix Calculation of Low-Energy Electron Collisions with Phosphoric Acid. *J. Phys. B At. Mol. Opt. Phys.* **2010**, *43*, 235203.

(42) Winstead, C.; McKoy, V. Low-Energy Electron Collisions with Gas-Phase Uracil. *J. Chem. Phys.* **2006**, *125* (17), 174304–174304.

(43) Li, X.; Sevilla, M. D.; Sanche, L. Density Functional Theory Studies of Electron Interaction with DNA: Can Zero eV Electrons Induce Strand Breaks? *J. Am. Chem. Soc.* **2003**, *125* (45), 13668–13669.

(44) Verma, P.; Narayanan S J, J.; Dutta, A. K. Electron Attachment to DNA: The Protective Role of Amino Acids. *J. Phys. Chem. A* **2023**, *127* (10), 2215–2227.

(45) Verma, P.; Mukherjee, M.; Bhattacharya, D.; Haritan, I.; Dutta, A. K. Shape Resonance Induced Electron Attachment to Cytosine: The Effect of Aqueous Media. *J. Chem. Phys.* **2023**, *159* (21), 214303.

(46) Martin, F.; Burrow, P. D.; Cai, Z.; Cloutier, P.; Hunting, D.; Sanche, L. DNA Strand Breaks Induced by 0-4 eV Electrons: The Role of Shape Resonances. *Phys. Rev. Lett.* **2004**, *93* (6), 6–9.

(47) Moiseyev, N. *Non-Hermitian Quantum Mechanics*; Cambridge University Press, 2011.

(48) Rescigno, T. N.; McCurdy, C. W.; Orel, A. E. Extensions of the Complex-Coordinate Method to the Study of Resonances in Many-Electron Systems. *Phys. Rev. A* **1978**, *17*, 1931–1938.

(49) Hernández Vera, M.; Jagau, T.-C. Resolution-of-the-Identity Second-Order Møller–Plesset Perturbation Theory with Complex Basis Functions: Benchmark Calculations and Applications to Strong-Field Ionization of Polyacenes. *J. Chem. Phys.* **2020**, *152* (17), 174103.

(50) White, A. F.; Epifanovsky, E.; McCurdy, C. W.; Head-Gordon, M. Second Order Møller-Plesset and Coupled Cluster Singles and Doubles Methods with Complex Basis Functions for Resonances in Electron-Molecule Scattering. *J. Chem. Phys.* **2017**, *146* (23), 234107.

(51) Moiseyev, N.; Corcoran, C. Autoionizing States of H2 and H2- Using the Complex-Scaling Method. *Phys Rev A* **1979**, *20* (3), 814–817.

(52) Jagau, T. C.; Bravaya, K. B.; Krylov, A. I. Extending Quantum Chemistry of Bound States to Electronic Resonances. *Annu. Rev. Phys. Chem.* **2017**, *68*, 525–553.

(53) Kanazawa, Y.; Ehara, M.; Sommerfeld, T. Low-Lying Π* Resonances of Standard and Rare DNA and RNA Bases Studied by the Projected CAP/SAC-CI Method. *J. Phys. Chem. A* **2016**, *120* (9), 1545–1553.

(54) Sajeev, Y.; Ghosh, A.; Vaval, N.; Pal, S. Coupled Cluster Methods for Autoionisation Resonances. *Int. Rev. Phys. Chem.* **2014**, *33* (3), 397–425.

(55) Jagau, T.-C.; Zuev, D.; Bravaya, K. B.; Epifanovsky, E.; Krylov, A. I. A Fresh Look at Resonances and Complex Absorbing Potentials: Density Matrix-Based Approach. *J. Phys. Chem. Lett.* **2014**, *5* (2), 310–315.

(56) Sajeev, Y.; Moiseyev, N. Reflection-Free Complex Absorbing Potential for Electronic Structure Calculations: Feshbach-Type Autoionization Resonances of Molecules. *J. Chem. Phys.* **2007**, *127* (3), 034105.

(57) Sommerfeld, T.; Ehara, M. Complex Absorbing Potentials with Voronoi Isosurfaces Wrapping Perfectly around Molecules. *J. Chem. Theory Comput.* **2015**, *11* (10), 4627–4633.

(58) Gayvert, J. R.; Bravaya, K. B. Projected CAP-EOM-CCSD Method for Electronic Resonances. *J. Chem. Phys.* **2022**, *156* (9), 094108.

(59) Gayvert, J. R.; Bravaya, K. B. Application of Box and Voronoi CAPs for Metastable Electronic States in Molecular Clusters. *J. Phys. Chem. A* **2022**, *126* (30), 5070–5078.

(60) Das, S.; Samanta, K. Recent Advances in the Study of Negative-Ion Resonances Using Multiconfigurational Propagator and a Complex Absorbing Potential. *ChemPhysChem* **2023**, *24* (3), e202200546.

(61) Hazi, A. U.; Taylor, H. S. Stabilization Method of Calculating Resonance Energies: Model Problem. *Phys Rev A* **1970**, *1* (4), 1109–1120.

(62) Landau, A.; Haritan, I.; Moiseyev, N. The RVP Method—From Real Ab-Initio Calculations to Complex Energies and Transition Dipoles. *Front. Phys.* **2022**, *10*.

(63) Landau, A.; Haritan, I.; Kaprálová-Žďánská, P. R.; Moiseyev, N. Atomic and Molecular Complex Resonances from Real Eigenvalues Using Standard (Hermitian) Electronic Structure Calculations. *J. Phys. Chem. A* **2016**, *120* (19), 3098–3108.

(64) Haritan, I.; Moiseyev, N. On the Calculation of Resonances by Analytic Continuation of Eigenvalues from the Stabilization Graph. *J. Chem. Phys.* **2017**, *147* (1), 014101.

(65) Abdoul-Carime, H.; Gohlke, S.; Fischbach, E.; Scheike, J.; Illenberger, E. Thymine Excision from DNA by Subexcitation Electrons. *Chem. Phys. Lett.* **2004**, *387* (4–6), 267–270.

(66) Burrow, P. D.; Gallup, G. A.; Scheer, A. M.; Denifl, S.; Ptasinska, S.; Märk, T.; Scheier, P. Vibrational Feshbach Resonances in Uracil and Thymine. *J. Chem. Phys.* **2006**, *124* (12), 124310–124310.

(67) Denifl, S.; Ptasińska, S.; Probst, M.; Hrušák, J.; Scheier, P.; Märk, T. D. Electron Attachment to the Gas-Phase DNA Bases Cytosine and Thymine. *J. Phys. Chem. A* **2004**, *108* (31), 6562–6569.

(68) Dora, A.; Bryjko, L.; Mourik, T. van; Tennyson, J. R-Matrix Study of Elastic and Inelastic Electron Collisions with Cytosine and Thymine. *J. Phys. B At. Mol. Opt. Phys.* **2012**, *45* (17), 175203.

(69) Berdys, J.; Skurski, P.; Simons, J. Damage to Model DNA Fragments by 0.25-1.0 eV Electrons Attached to a Thymine Π* Orbital. **2004**.

(70) McAllister, M.; Kazemigazestane, N.; Henry, L. T.; Gu, B.; Fabrikant, I.; Tribello, G. A.; Kohanoff, J. Solvation Effects on Dissociative Electron Attachment to Thymine. *J. Phys. Chem. B* **2019**, *123* (7), 1537–1544.

(71) Bouskila, G.; Landau, A.; Haritan, I.; Moiseyev, N.; Bhattacharya, D. Complex Energies and Transition Dipoles for Shape-Type Resonances of Uracil Anion from Stabilization Curves via Padé. *J. Chem. Phys.* **2022**, *156* (19), 194101.

(72) Cheng, H.-Y.; Chen, C.-W. Energy and Lifetime of Temporary Anion States of Uracil by Stabilization Method. *J. Phys. Chem. A* **2011**, *115* (35), 10113–10121.

(73) Cooper, G.; Clarke, C.; Verlet, J. Electron Impact Resonances of Uracil in an Aqueous Environment from Anion Photoelectron Imaging. *J. Phys. B At. Mol. Opt. Phys.* **2023**, *56*.

(74) Verma, P.; Ghosh, D.; Dutta, A. K. Electron Attachment to Cytosine: The Role of Water. *J. Phys. Chem. A* **2021**, *125* (22), 4683–4694.

(75) Bhaskaran, R.; Sarma, M. Low Energy Electron Induced Cytosine Base Release in 2′-Deoxycytidine-3′-Monophosphate via Glycosidic Bond Cleavage: A Time-Dependent Wavepacket Study. *J. Chem. Phys.* **2014**, *141* (10), 104309–104309.

(76) Arora, S.; Narayanan, S. J. J.; Dutta, A. K. How Good Is the Time-Dependent DFT Method for Simulating Anionic Shape Resonances of DNA Nucleobases? *J. Chem. Sci.* **2025**, *137* (4), 119.

(77) Tripathi, D.; Pyla, M.; Dutta, A. K.; Matsika, S. Impact of Solvation on the Electronic Resonances in Uracil. *Phys. Chem. Chem. Phys.* **2025**, *27* (7), 3588–3601.

(78) Pyla, M.; Matsika, S. Partial Widths of Shape Resonances in Pyridine and Uracil Using the Stabilization Method. *J. Chem. Phys.* **2026**, *164* (1), 014308.

(79) Narayanan S J, J.; Tripathi, D.; Haritan, I.; Dutta, A. K. The Effect of Aqueous Medium on Nucleobase Shape Resonances: Insights from Microsolvation. *J. Phys. Chem. A* **2025**, *129* (48), 11179–11188.

(80) Colson, A. O.; Besler, B.; Sevilla, M. D. Ab Initio Molecular Orbital Calculations on DNA Base Pair Radical Ions: Effect of Base Pairing on Proton-Transfer Energies, Electron Affinities, and Ionization Potentials. *J. Phys. Chem.* **1992**, *96* (24), 9787–9794.

(81) Al-Jihad, I.; Smets, J.; Adamowicz, L. Covalent Anion of the Canonical Adenine−Thymine Base Pair. Ab Initio Study. *J. Phys. Chem. A* **2000**, *104* (13), 2994–2998.

(82) Smets, J.; Jalbout, A. F.; Adamowicz, L. Anions of the Hydrogen-Bonded Guanine–Cytosine Dimer – Theoretical Study. *Chem. Phys. Lett.* **2001**, *342* (3), 342–346.

(83) Riplinger, C.; Neese, F. An Efficient and near Linear Scaling Pair Natural Orbital Based Local Coupled Cluster Method. *J. Chem. Phys.* **2013**, *138* (3), 034106–034106.

(84) Dutta, A. K.; Saitow, M.; Demoulin, B.; Neese, F.; Izsák, R. A Domain-Based Local Pair Natural Orbital Implementation of the Equation of Motion Coupled Cluster Method for Electron Attached States. *J. Chem. Phys.* **2019**, *150* (16), 164123.

(85) Landau, A.; Haritan, I. The Clusterization Technique: A Systematic Search for the Resonance Energies Obtained via Padé. *J. Phys. Chem. A* **2019**, *123* (24), 5091–5105.

(86) Weigend, F.; Häser, M. RI-MP2: First Derivatives and Global Consistency. *Theor. Chem. Acc.* **1997**, *97* (1), 331–340.

(87) Weigend, F.; Häser, M.; Patzelt, H.; Ahlrichs, R. RI-MP2: Optimized Auxiliary Basis Sets and Demonstration of Efficiency. *Chem. Phys. Lett.* **1998**, *294* (1), 143–152.

(88) Weigend, F.; Ahlrichs, R. Balanced Basis Sets of Split Valence, Triple Zeta Valence and Quadruple Zeta Valence Quality for H to Rn: Design and Assessment of Accuracy. *Phys. Chem. Chem. Phys.* **2005**, *7* (18), 3297–3305.

(89) Pracht, P.; Bohle, F.; Grimme, S. Automated Exploration of the Low-Energy Chemical Space with Fast Quantum Chemical Methods. *Phys Chem Chem Phys* **2020**, *22* (14), 7169–7192.

(90) Tripathi, D.; Dutta, A. K. Electron Attachment to DNA Base Pairs: An Interplay of Dipole- And Valence-Bound States. *J. Phys. Chem. A* **2019**, *123* (46), 10131–10138.

(91) Mukherjee, M.; Tripathi, D.; Dutta, A. K. Water Mediated Electron Attachment to Nucleobases: Surface-Bound vs Bulk Solvated Electrons. *J. Chem. Phys.* **2020**, *153* (4), 044305.

(92) Dunning, T. H. Gaussian Basis Sets for Use in Correlated Molecular Calculations. I. The Atoms Boron through Neon and Hydrogen. *J. Chem. Phys.* **1989**, *90* (2), 1007–1023.

(93) Stoychev, G. L.; Auer, A. A.; Neese, F. Automatic Generation of Auxiliary Basis Sets. *J. Chem. Theory Comput.* **2017**, *13* (2), 554–562.

(94) Neese, F. The ORCA Program System. *Wiley Interdiscip. Rev. Comput. Mol. Sci.* **2012**, *2* (1), 73–78.

(95) Neese, F. Software Update: The ORCA Program System—Version 5.0. *WIREs Comput. Mol. Sci.* **2022**, *12* (5), e1606.

(96) Safrai, Y.; Haritan, I. Automatic-Rvp: RVP Program, 2021. https://pypi.org/project/automatic-rvp/ (accessed 2023-08-03).

(97) Arora, S.; Narayanan S J, J.; Haritan, I.; Adhikary, A.; Dutta, A. K. Effect of Protein Environment on the Shape Resonances of RNA Pyrimidine Nucleobases: Insights from a Model System. *J. Chem. Phys.* **2025**, *163* (13), 134103.

(98) Aflatooni, K.; Gallup, G. A.; Burrow, P. D. Electron Attachment Energies of the DNA Bases. *J. Phys. Chem. A* **1998**, *102* (31), 6205–6207.

(99) Tonzani, S.; Greene, C. H. Low-Energy Electron Scattering from DNA and RNA Bases: Shape Resonances and Radiation Damage. *J. Chem. Phys.* **2006**, *124* (5), 054312.

(100) Dutta, A. K.; Sengupta, T.; Vaval, N.; Pal, S. Electron Attachment to DNA and RNA Nucleobases: An EOMCC Investigation. *Int. J. Quantum Chem.* **2015**, *115* (12), 753–764.

(101) Tripathi, D.; Dutta, A. K. Bound Anionic States of DNA and RNA Nucleobases: An EOM-CCSD Investigation. *Int. J. Quantum Chem.* **2019**, *119* (9), e25875.

(102) J, J. N. S.; Tripathi, D.; Haritan, I.; Adhikary, A.; Pandey, B.; Dutta, A. K. The Effect of Base-Pairing on the Shape Resonances of Nucleobases. *arXiv [physics.chem-ph]*, 2026.

(103) Li, Z.; Cloutier, P.; Sanche, L.; Wagner, J. R. Low-Energy Electron-Induced DNA Damage: Effect of Base Sequence in Oligonucleotide Trimers. *J. Am. Chem. Soc.* **2010**, *132* (15), 5422–5427.

(104) Winstead, C.; McKoy, V.; d'Almeida Sanchez, S. Interaction of Low-Energy Electrons with the Pyrimidine Bases and Nucleosides of DNA. *J. Chem. Phys.* **2007**, *127* (8), 085105.

(105) Dora, A.; Bryjko, L.; van Mourik, T.; Tennyson, J. Low-Energy Electron Scattering with the Purine Bases of DNA/RNA Using the R-Matrix Method. *J. Chem. Phys.* **2012**, *136* (2), 024324.

(106) Winstead, C.; McKoy, V. Interaction of Low-Energy Electrons with the Purine Bases, Nucleosides, and Nucleotides of DNA. *J. Chem. Phys.* **2006**, *125* (24), 244302.

## Tables and figures:

**Table 1:** *The comparison of resonance positions and widths (in parentheses) of* ***Thymine*** *with previous experimental and theoretical results. All values are in eV.*

| Thymine | Er(Γ) (eV) | | |
|---|---|---|---|
| | **1π*** | **2π*** | **3π*** |
| **RVP/EA-EOM-DLPNO-CCSD (This work)** | **0.69(0.014)** | **2.35(0.028)** | **5.69(0.102)** |
| **GPA/EOM-EA-CCSD**[36] | 0.68(0.02) | 2.32(0.073) | 5.02(0.58) |
| **CAP/SAC-CI**[53] | 0.67(0.11) | 2.28(0.15) | 5.14(0.41) |
| **R-matrix/u-CC**[68] | 0.53(0.08) | 2.41(0.10) | 5.26 |
| **R-matrix/SEP**[68] | 0.60(0.11) | 2.73(0.11) | 5.52(0.57) |
| **SMC/SEP**[104] | 0.30 | 0.19 | 5.70 |
| **R-matrix/SE**[99] | 2.40(0.20) | 5.50(0.60) | 7.90(1.00) |
| **Expt.**[98] | 0.29 | 1.71 | 4.05 |

**Table 2:** *The comparison of resonance energies and widths (in parentheses) of* ***Adenine*** *with previous experimental and theoretical results. Values are given in eV.*

| Adenine | Er(Γ) (eV) | | | |
|---|---|---|---|---|
| | **1π*** | **2π*** | **3π*** | **4π*** |
| **RVP/EA-EOM-DLPNO-CCSD (This work)** | **1.11(0.013)** | **1.98(0.026)** | **2.97(0.038)** | **7.20(0.204)** |
| **GPA/EOM-EA-CCSD**[36] | 1.13(0.04) | 2.01(0.05) | 2.96(0.14) | 6.78(0.21) |
| **CAP/SAC-CI**[53] | 0.89(0.13) | 1.93(0.29) | 2.72(0.17) | 6.65(0.57) |
| **R-matrix/u-CC**[105] | 1.58(0.22) | 2.44(0.14) | 4.38(0.67) | 7.94(0.57) |
| **R-matrix/SEP**[105] | 1.30(0.14) | 2.12(0.09) | 3.12(0.28) | 7.07(0.24) |
| **SMC/SEP**[106] | 1.10 | 1.80 | 4.10 | |
| **R-matrix/SE**[99] | 2.40(0.20) | 3.20(0.20) | 4.40(0.30) | |
| **Expt.**[98] | 0.54 | 1.36 | 2.17 | |

**Table 3:** *The comparison of resonance energies and widths of* ***Adenine*** *and* ***Thymine*** *with the* ***AT*** *base pair (linear). (The color coding identifies whether an AT resonance is adenine- or thymine-centered, with each assigned the same color as the corresponding isolated nucleobase resonance.)*

| System | 1π* | 2π* | 3π* | 4π* | 5π* | 6π* | 7π* |
|---|---|---|---|---|---|---|---|
| AT | **0.67 (0.011)** | **1.26 (0.016)** | **1.93 (0.020)** | **2.43 (0.035)** | **3.16 (0.044)** | **5.67 (0.084)** | **6.80 (0.116)** |
| Adenine | **1.11 (0.013)** | **1.98 (0.026)** | **2.97 (0.038)** | **7.20 (0.204)** | | | |
| Thymine | **0.69 (0.014)** | **2.35 (0.028)** | **5.71 (0.102)** | | | | |

**Table 4:** *The comparison of resonance parameters for AT base pair in the linear and stacked geometry.*

| | | | | $E_R$ (Γ) in eV | | | |
|---|---|---|---|---|---|---|---|
| **System** | 1π* | 2π* | 3π* | 4π* | 5π* | 6π* | 7π* |
| **AT (linear)** | 0.67 (0.011) | 1.26 (0.016) | 1.93 (0.020) | 2.43 (0.035) | 3.16 (0.044) | 5.67 (0.084) | 6.80 (0.116) |
| **AT (stacked)** | 0.64 (0.009) | 0.88 (0.014) | 1.65 (0.016) | 2.23 (0.025) | 2.73 (0.035) | 5.33 (0.071) | 6.51 (0.108) |

**Table 5:** *The resonance positions and widths of* ***AA*** *and* ***TT*** *stacked base pairs.*

| System | 1π* | 2π* | 3π* | 4π* | 5π* | 6π* | 7π* | 8π* |
|---|---|---|---|---|---|---|---|---|
| **sAA** | **1.05 (0.012)** | **1.17 (0.018)** | **1.92 (0.021)** | **2.03 (0.033)** | **2.59 (0.037)** | **3.16 (0.057)** | **6.77 (0.127)** | **7.72 (0.228)** |
| **Adenine** | **1.11 (0.013)** | **1.98 (0.026)** | **2.97 (0.038)** | **7.20 (0.204)** | | | | |
| **sTT** | **0.61 (0.010)** | **0.73 (0.019)** | **1.98 (0.024)** | **2.43 (0.033)** | **5.42 (0.066)** | **5.73 (0.110)** | | |
| **Thymine** | **0.69 (0.014)** | **2.35 (0.028)** | **5.69 (0.102)** | | | | | |

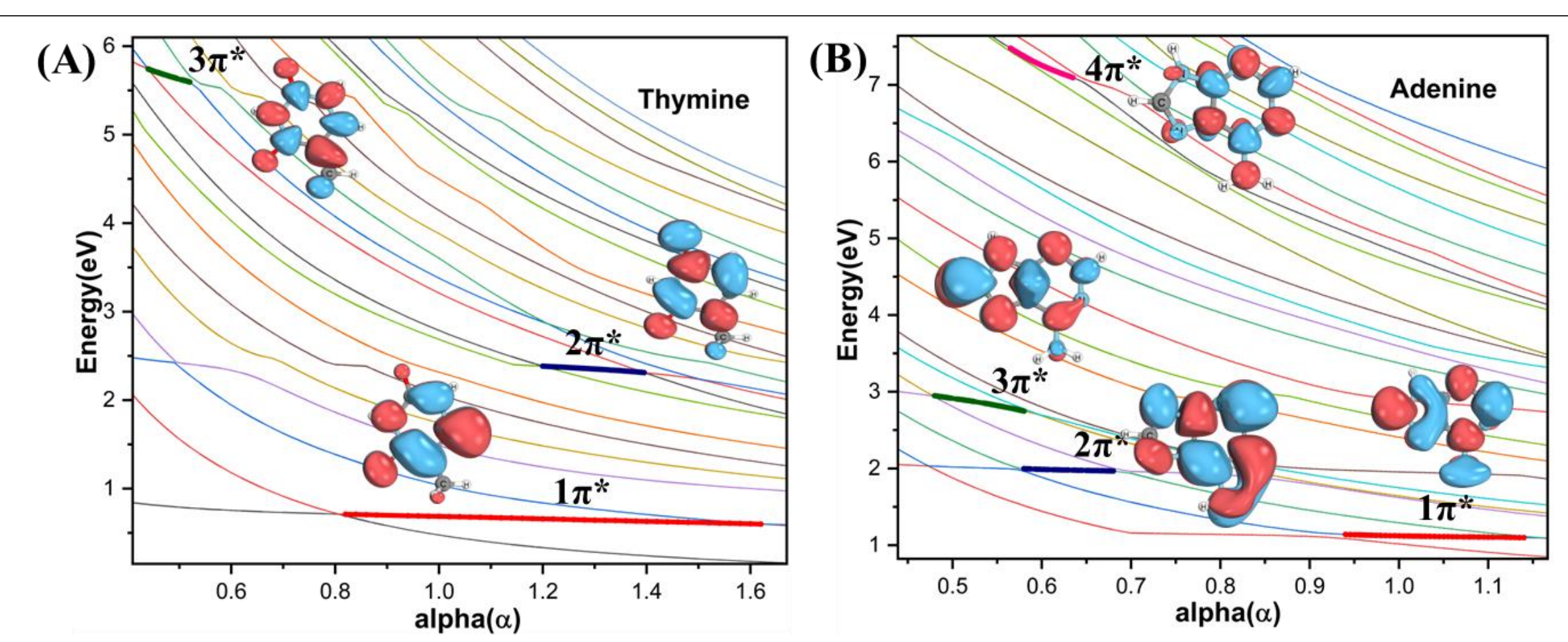

**Figure 1:** *Stabilization plots of (A) Thymine (B) Adenine with the highlighted stable regions and natural orbitals corresponding to the shape resonances.*

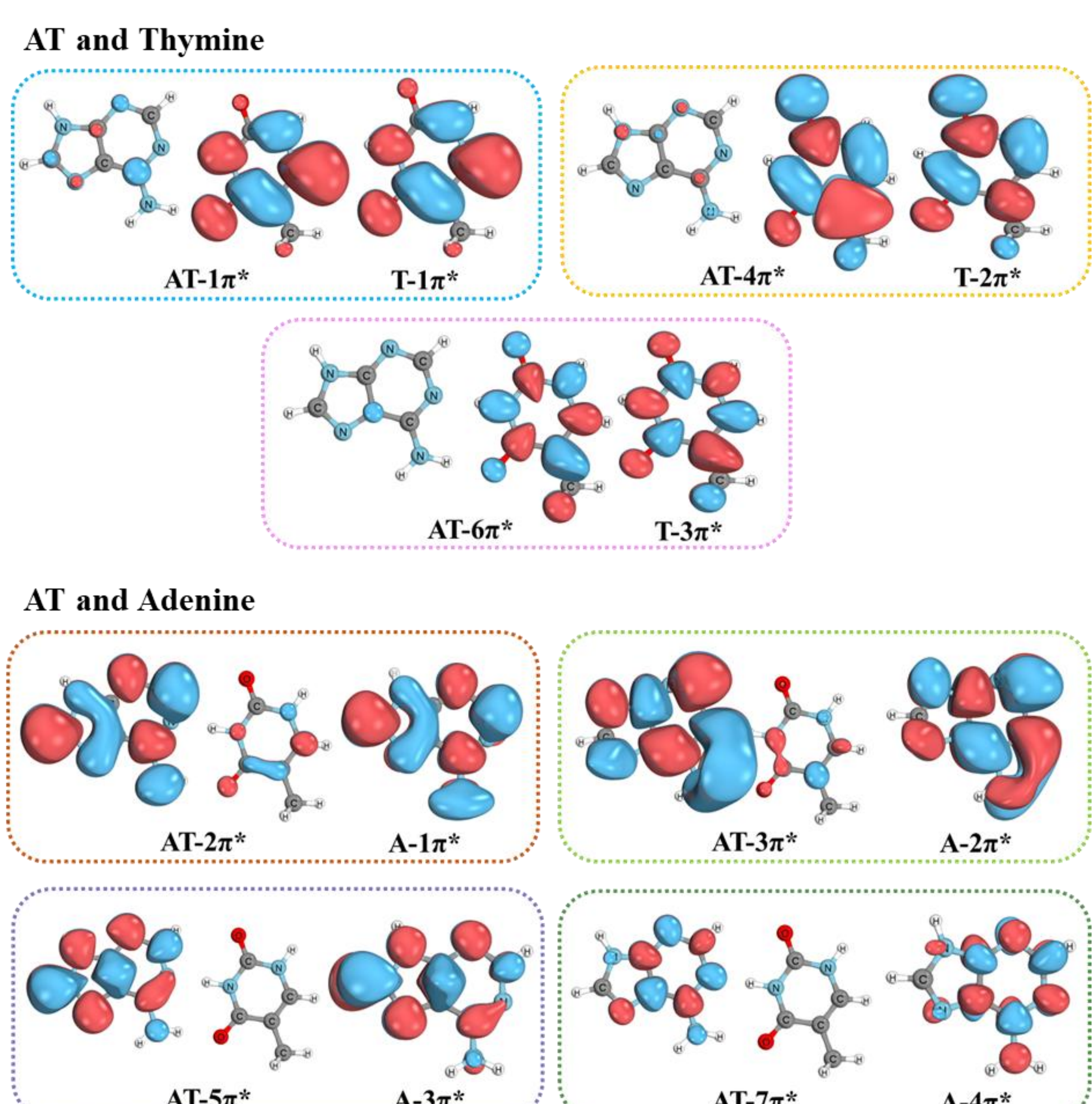

**Figure 2:** *Natural orbitals comparison analysis of adenine and thymine with the AT base pair.*

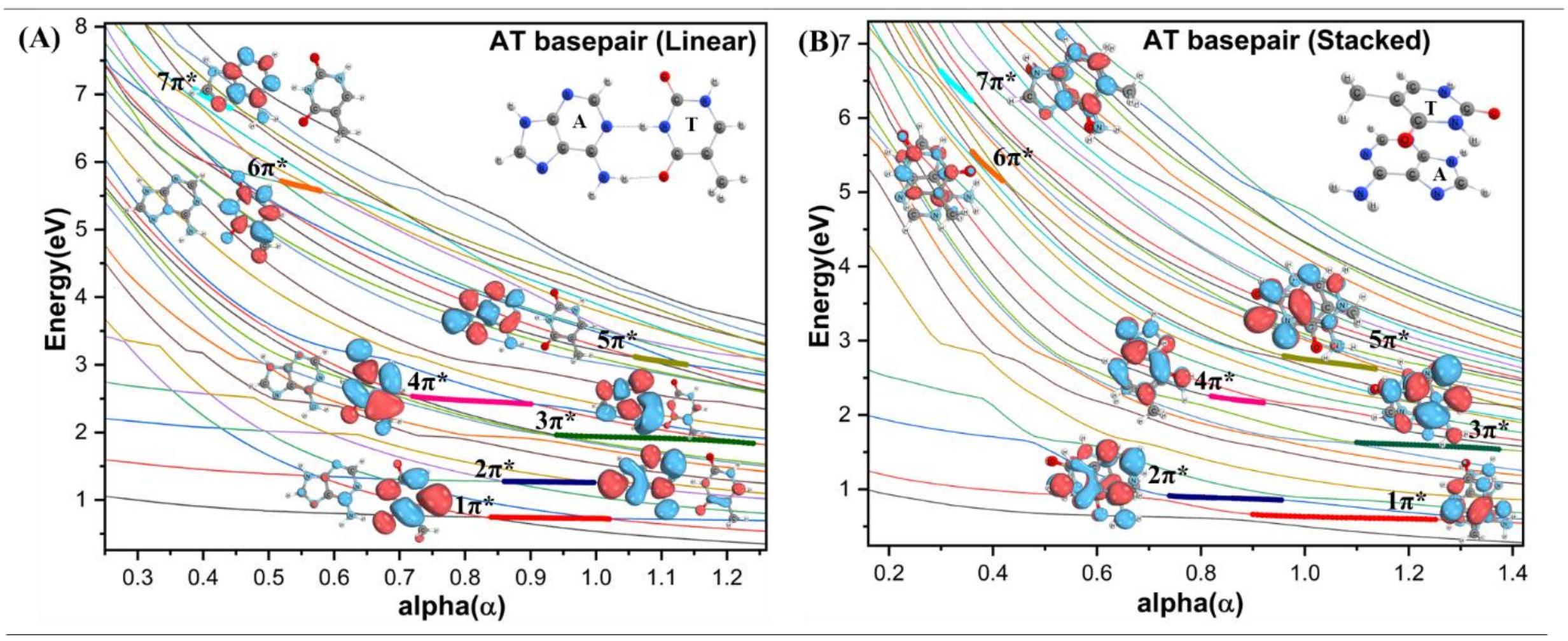

**Figure 3:** *Energy stabilization plots and natural orbitals for(A) linear and (B) stacked AT base pairs.*

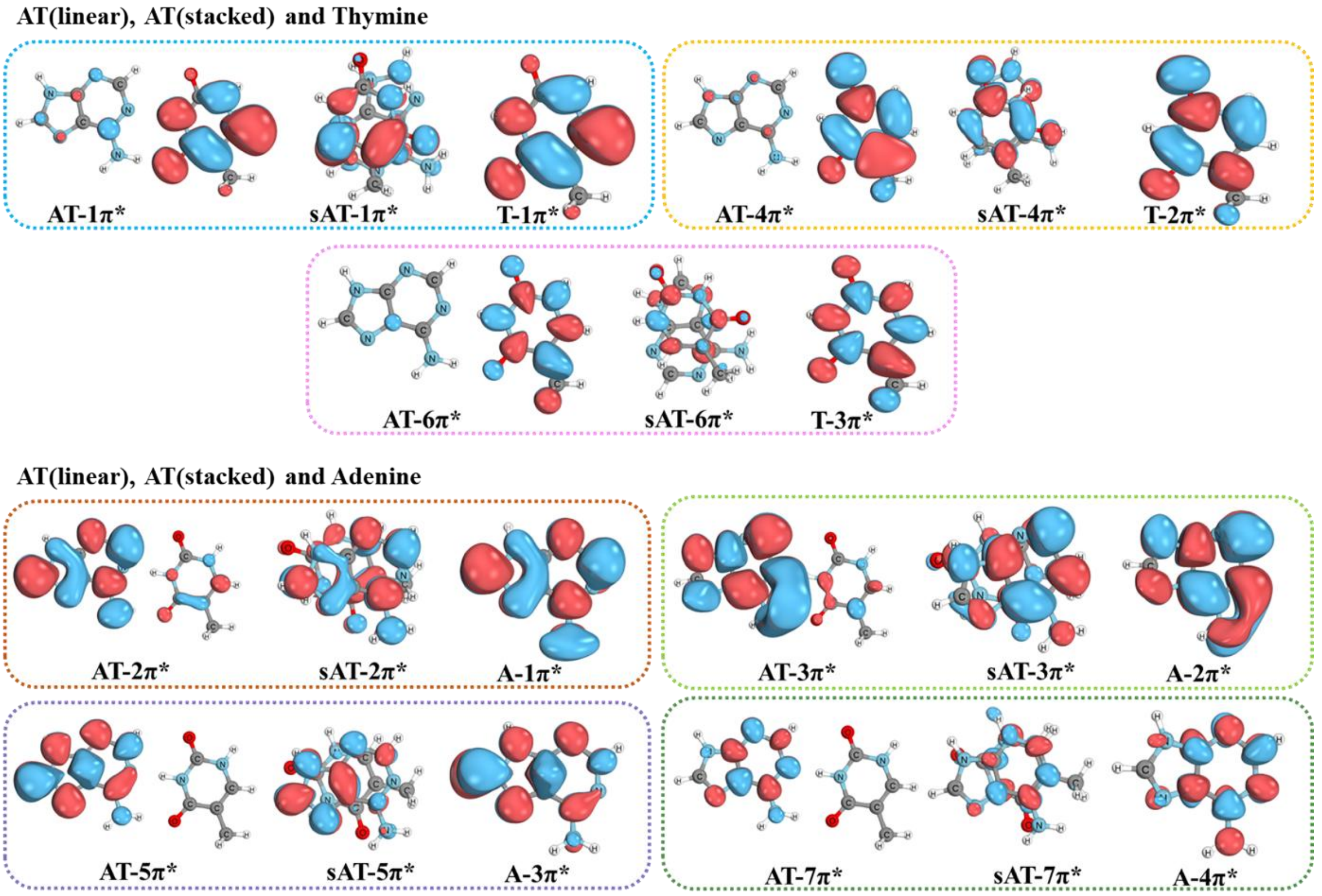

**Figure 4:** *Natural orbitals comparison analysis of isolated adenine and thymine nucleobases with the AT base pair (linear and stacked conformations).*

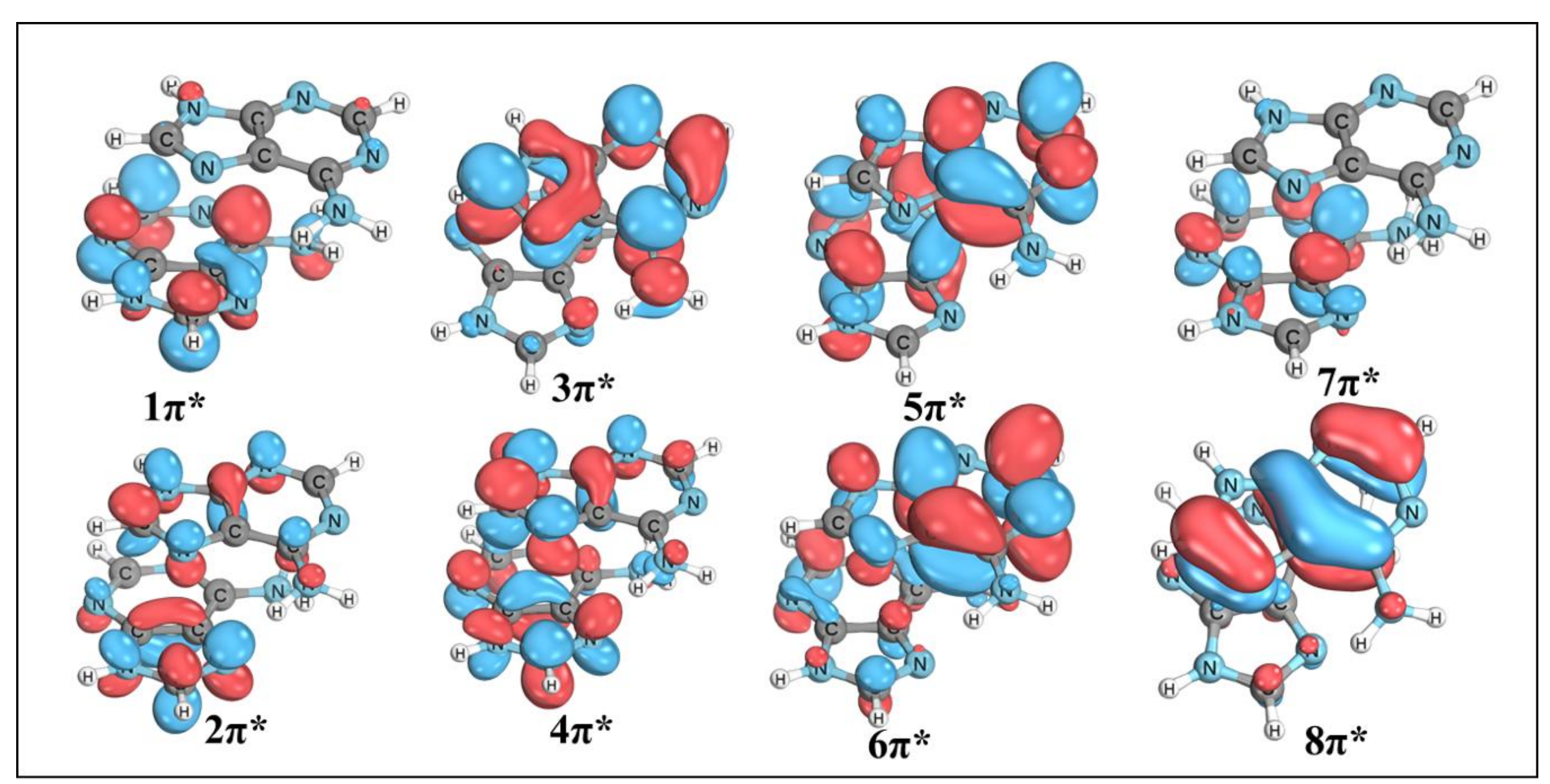

**Figure 5:** *Natural orbitals corresponding to adenine-adenine stacked base pair.*

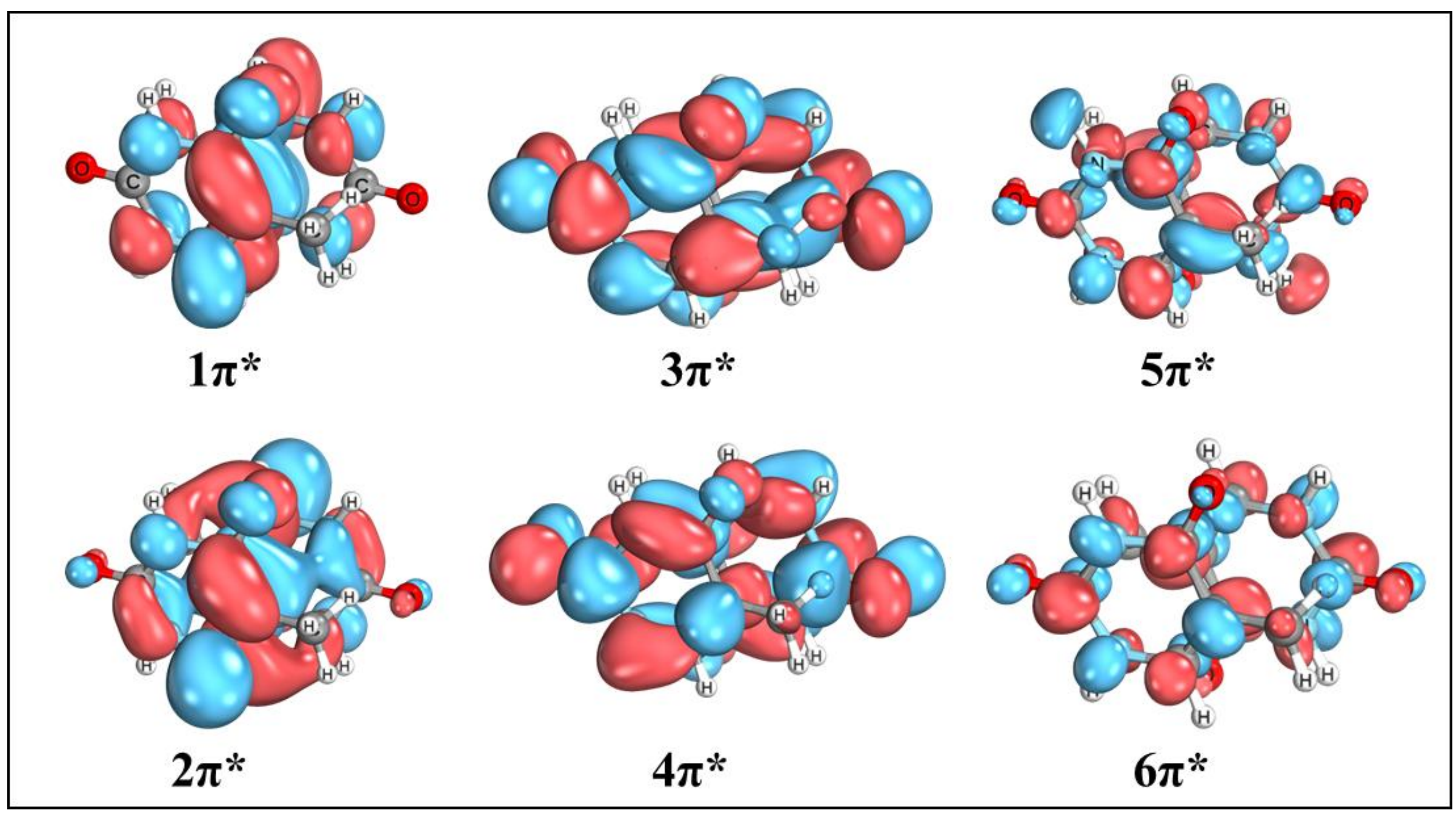

**Figure 6:** *Natural orbitals corresponding to the thymine-thymine stacked base pair.*